
\magnification=1200
\baselineskip=18truept

\def\draftversion{N}

\if \draftversion Y


\fi

\def\chiral{{\rm{\bf C}}}
\def\wilson#1{{\rm{\bf B^#1}}}
\def\ham#1{{\rm{\bf H^#1}}}

\line{\hfill IASSNS--HEP--95/19}
\line{\hfill RU--95--16}
\line{\hfill CU--TP--680}
\vskip 2truecm
\centerline{\bf A simulation of the Schwinger model in
the overlap formalism}

\vskip 1truecm
\centerline{Rajamani Narayanan$^{a}$, Herbert Neuberger$^{b}$
 and Pavlos Vranas$^c$}
\vskip 0.5truecm
\centerline{${}^a$\it School of Natural Sciences,
Institute for Advanced Study,}
\centerline{\it Olden Lane, Princeton, NJ 08540}
\vskip .5truecm
\centerline{${}^b$\it Department of Physics and Astronomy,
Rutgers University }
\centerline{\it Piscataway, NJ 08855-0849}
\vskip .5truecm
\centerline{${}^c$\it Department of Physics, Columbia University, New York,
NY 10027}

\vfill
\centerline{\bf Abstract}
\vskip 0.75truecm
In the continuum, the single flavor massless Schwinger
model has an exact global axial $U(1)$ symmetry in the
sector of perturbative gauge fields. This symmetry is
explicitly broken by gauge fields with nonzero
topological charge inducing a nonzero
expectation value
for the bilinear $\bar\psi\psi$. We show that a lattice formulation
of this model, using the overlap formalism to treat the massless fermions,
explicitly exhibits this phenomenon.
A Monte Carlo simulation of the complete system
yields the correct value of the fermion condensate and
shows  unambiguously that it originates
from the sector of topological charge equal to
unity.
\vfill
\eject

\centerline{\bf Introduction}
\bigskip
In chiral gauge theories fermions appear
bilinearly in the action
and interact with each other only by the exchange of gauge
bosons.
The bilinearity in fermion
fields is preserved in most methods of regularization.
Then the regularization becomes defined for any
single (complex) irreducible representation carried by the chiral fermions,
be it anomalous or not. Using this definition one can construct, in
particular, vector gauge theories where the multiplets are
assembled in left--handed/right--handed pairs per representation.
Massless QCD is an example of such a vector gauge theory
and it has an anomalous
global $U(1)$ symmetry that is violated by topology
induced 't Hooft vertices [1]. Therefore, any regularization
of chiral fermions that preserves bilinearity should
break the anomalous $U(1)$, but conserve the nonanomalous global
symmetries. Some of the existing proposals may have
problems in this respect; we know of no examples where this point
has been tested with quantum fluctuations of gauge bosons taken
into account.

Here we subject the overlap formalism to this test.
For smooth, classical,
gauge fields we already know that the overlap formalism works correctly [2,3].
The purpose of this paper is to report a Monte Carlo simulation
of the {\it single flavor, strictly massless} Schwinger model [4]
formulated on the lattice.
Each of the two
chiral components of the Dirac fermion
are realized using the overlap formula. Of course, this model
can also be formulated on the lattice
 using conventional techniques [5], but then a
massless limit has to be taken, there is difficulty in realizing a single
flavor, there
is no factorization of the theory into left and right sectors for
fixed gauge fields,
 and the role of topology is obscured. Anyhow, our main
point is not to present yet another solution of the
Schwinger model, but, to establish a useful test for proposals
to regularize chiral gauge theories and show that the overlap
passes this test successfully.

In order to make it absolutely clear that the overlap
succeeds in computing something that  no
previous approach to regularizing chiral fermions could,
 we picked as our
observable just the 't Hooft vertex $\bar\psi_R \psi_L$ which
together with its hermitian conjugate makes up the
mass bilinear $\bar\psi\psi$ more familiar in this context [6]. Our
observable receives contributions only from gauge fields of
topological charge one. The Schwinger model is a good model to pick
because it has the necessary physics ingredients, its study
is not particular demanding computationally and, above all,
it is exactly soluble in the continuum, even for finite size systems.
Various aspects of a two flavor version of the model have been investigated
in detail in recent years both in the continuum and on the lattice
by Azakov, Joos and Dilger [7]. Here we rely mostly on the
continuum analysis of Sachs and Wipf [6].

\bigskip\bigskip
\centerline{\bf The model}
\bigskip
The continuum action
in Euclidean space is:
$$
S[\bar\psi,\psi,A_\mu] =
{1\over{2g^2}} \int (\epsilon_{\mu\nu}\partial_\mu A_\nu )^2 -
\int \bar\psi_L \sigma_\mu (\partial_\mu + i A_\mu )\psi_L -
\int \bar\psi_R \sigma^*_\mu (\partial_\mu +i A_\mu )\psi_R
$$
where $\mu ,\nu =1,2$, $\sigma_1=1$ and $\sigma_2=i$.
$\bar\psi_{R,L}, \psi_{R,L}$ are Grassmann variables and $A_\mu$ is
a $U(1)$ gauge field. The model is defined on a torus of fixed
physical size
$l\times l$ with
periodic boundary conditions for the electric field,
$E=\epsilon_{\mu\nu}\partial_\mu A_\nu $,
and the fermions. The topological charge $q$ is an integer given
by $q={1\over{2\pi}}\int E$.
The expectation value of the fermion bilinear is
$$
<\bar\psi \psi > ={{\int [ d\bar\psi d\psi dA_\mu ]
( \bar\psi_R (0) \psi_L (0) + \bar\psi_L (0) \psi_R (0))
e^{-S(\bar\psi, \psi, A_\mu)}
}\over{\int [ d\bar\psi d\psi dA_\mu ]
e^{-S(\bar\psi, \psi, A_\mu)}}}\equiv {N\over D}$$

The lattice formulation begins by
replacing the continuum torus with a toroidal square
lattice consisting of $L$ sites in both directions. The
lattice spacing $a$ is $a=l/L$.
Gauge fields $U_\mu (x)  =e^{iaA_\mu (x)}$
are associated with the links of the lattice.
The overlap formalism [2] is used to deal with the chiral fermions
on the lattice. This results in the following
expressions for $N$ and $D$ appearing in the expectation value
of the fermion bilinear:
$$
N={1\over L^2} \sum_{x\alpha}
\int \prod dU_\mu (x) e^{-S_g (U)}
\Bigl[
 |{}_U\!\! <L-|a_{x\alpha}|L+>_{U}\!|^2 +
 |{}_U\!\! <L-|a^\dagger_{x\alpha}|L+>_{U}\!|^2
\bigr]
$$
$$
D=\int \prod dU_\mu (x) e^{-S_g (U)} |{}_U\!\! <L-|L+>_{U}\!|^2
$$
The $a$--term in $N$ comes from gauge fields carrying
topological charge $+1$
and is the overlap expression for $<\bar\psi_R \psi_L >$
while the $a^\dagger$--term comes from charge $-1$
and represents $<\bar\psi_L \psi_R >$. In our simulations
we could have looked at only one of these terms; we took
both in order to increase statistics. Analytically, the two
overlap matrix elements have exactly equal magnitudes.
The CP violating angle $\theta$ is set to zero and therefore
only the square of the absolute value appears in the above formula.
The states $|L\pm>_U$ are the ground states of many body Hamiltonians
$$
{\cal H}^\pm =\sum_{x\alpha,y\beta} a^\dagger_{x,\alpha }
\ham\pm(x\alpha,y\beta; U)
a_{y,\beta }~~~~~~~~~
\{a^\dagger_{x,\alpha }, a_{y,\beta }\} =\delta_{\alpha,\beta}\delta_{x,y}.$$
$\alpha,\beta=1,2$ and $x=(x_1 , x_2 )$ with
$x_\mu =0,1,2,.....,L -1 $. The single particle hermitian
hamiltonians $\ham\pm$ are given by:
$$\eqalign{
\ham\pm & =\pmatrix {\wilson\pm &\chiral\cr \chiral^\dagger&-\wilson\pm};\cr
\chiral(x,y) & ={1\over 2}
\sum_\mu \sigma_\mu (\delta_{y,x+\mu}U_\mu (x)-
\delta_{x,y+\mu}U_\mu^* (y) );\cr
\wilson\pm (x,y) & = {1\over 2} \sum_\mu
(2\delta_{x,y} - \delta_{y,x+\mu}U_\mu(x) - \delta_{x,y+\mu}U_\mu^* (y))
\pm m\delta_{x,y}\cr}
$$
The parameter $m$ can be chosen in the range $0 < m < 1$ and is set to $m=0.9$
here.
Note that topological charge $\pm 1$ means that the number of
single particle states filled in $|L+>_U$ and $|L->_U$ differ
by $\pm 1$.

We need a gauge action and an associated Monte Carlo method
that will give reasonable estimates for the topological charge
distribution. For this reason, we would like to avoid correlations
between configurations as far as possible. With this in mind,
we pick a single plaquette action,
$$S_g(U) = \sum_p s_g(U_p),$$
for the gauge action appearing in the formulae for $N$ and $D$.
$U_p$ is the oriented product of the four link elements making up the plaquette
$p$.
For such an action the plaquettes are almost independent variables
in the pure gauge theory.
To minimize the finite lattice spacing effects
arising from the gauge action, we chose
the ``heat-kernel'' action [8]:
$$
e^{-s_g (U_p )}=\sum_{m\in {\bf Z}} e^{-{{g^2 m^2}\over 2}} [U_p]^m$$
This choice assures exact scaling of the string
tension with the coupling $g$ in the pure gauge theory.

\bigskip\bigskip
\centerline{\bf The Monte Carlo method}
\bigskip
To compute the bilinear,
we use a Monte Carlo method where configurations are drawn independently of
each other. This avoids correlations between sequential samplings.
To this end we
replace $N$ by $N^\prime$ and $D$ by $D^\prime$
with
$N^\prime={N\over Z}$ and $D^\prime={D\over Z}$ where
$$
Z=\int \prod dU_\mu (x) e^{- S^\prime_g (U )}$$
Gauge configurations will be generated
stochastically using the distribution $e^{- S^\prime_g (U )}$ and,
denoting averages w.r.t. $e^{- S^\prime_g (U )}$ by $<<..>>$, we evaluate
probabilistically
$$
N^\prime = {1\over{L^2}}
\sum_{x\alpha} <<e^{- S_g (U ) + S^\prime_g (U )}
\Bigl[
 |{}_U\!\!<L-|a_{x,\alpha} |L+>_{U}\!|^2 +
 |{}_U\!\!<L-|a^\dagger_{x,\alpha} |L+>_{U}\!|^2
\Bigr] >>
$$
$$D^\prime =<<e^{- S_g (U ) + S^\prime_g (U )}
|{}_U\!\!<L-|L+>_{U}\!|^2 >>.$$
$S_g^\prime (U)$ is picked to be
$$
S_g^\prime (U) = S_g (U) - s_g (U_{p^*})$$
where $p^*$ is some arbitrarily chosen plaquette.
The probability
distribution $e^{S^\prime_g (U )}$ factorizes into one factor of
$e^{-s_g (U_p )}$ for each $p\ne p^*$.
To generate a configuration, one draws a set of $(L^2-1)$  $U_p$ variables,
each according to the plaquette factor $e^{-s_g (U_p)}$.
This fixes $(L^2-1)$ out of the $(L^2 + 1)$ link variables remaining
after choosing a gauge on the $L\times L$ lattice.
Two more link variables determining
Polyakov loops winding around the torus are drawn randomly since
the gauge action is unaffected by these loops.
Since we are computing gauge invariant quantities the exact nature of
gauge fixing is irrelevant and there is no need to average over gauge
transformations.

\bigskip\bigskip
\centerline{\bf The parameters}
\bigskip
Having described the Monte Carlo method, we now proceed to determine
the parameters of the model.
Since we are treating the overlap factors as observables in a pure
gauge theory, this method will work only if the fluctuations in these
factors are not too rapid. Technically, they are
of the order of
$e^{cL^2}$ and since we are simulating a system of finite
{\it physical} size $l$, $c$ is also dependent on $L$.
Based on the exact solution in the continuum we choose $l$
so that the fluctuations in the fermionic determinant are small.
We feel confident in using the continuum results for our
estimates since we have already shown that the overlap agrees
with the continuum very well for sufficiently smooth fields [2,3].
The zero momentum mode of the electrical field is quantized by
topology, so the dependence on it is special. The fermionic
determinant also depends on the zero momentum component of the gauge fields,
but these are only two variables and the dependence though sizable
is not too violent. Previous work assures us that the overlap
will follow continuum behavior faithfully also for these variables [9].
There remain now the nonzero momentum modes of the electrical field
$\tilde E (k)\equiv \phi(k) k^2$. The bosonic action per mode
is $e^{-{1\over{2 g^2}}|\phi (k) |^2 k^4}$ yielding $ |\phi (k) |^2
\approx {{2g^2}\over {k^4}}$. The fermion determinant factorizes
among the modes [2] and, per mode, is given by
$e^{-{1\over{2\pi}}|\phi (k) |^2 k^2}
\approx e^{-{{\mu^2}\over{k^2}}}$ where $\mu={g\over \sqrt{\pi}}$.
The smallest nonzero momentum is $k={{2\pi}\over l}$ so we get a typical
contribution of the order $e^{-{{(\mu l )^2} \over {4\pi^2 }}}$.
For $\mu l \approx 3$, the typical number in the exponent is expected to
be quite small.
If we sum over the higher modes the result
changes a little  but we do not need to complete
this exercise for our simple estimates.

Apart from the overlap factor, we also have the factor
$e^{-S_g(U) + S^\prime_g(U)}=e^{-s_g (U_{p^*} )}$ with
$U^*_{p^*} =\prod_{p\ne p^* } U_p$.
We should note that our method does not implement importance sampling
for this factor. We were willing to accept
this drawback in exchange for the unhindered generation
of the topological charge distribution. Since our physical volume
is fixed,
the coupling becomes weaker
as the lattice gets finer and $e^{s_g (U_p )}$ gets more peaked. Its
sampling however has a roughly constant and quite broad width, so,
unlike in the usual case where statistical errors per lattice decrease
with size due to self averaging, here the statistical error will
stay more or less constant as a function of size and therefore
the computing cost per given accuracy will increase
dramatically with size. Although this can be remedied to some extent,
it is not necessary for the computation presented here.

Gauge
configurations contributing to $N^\prime$ have unit
lattice topological charge
and those that contribute to $D^\prime$ have zero lattice topological charge.
In all other cases the respective overlap vanishes identically. There
is a probability $p(0)\equiv p$ to generate gauge field configurations
of zero lattice topological charge.
In a sample of $n$ lattices
there will be $j\approx pn$ lattices contributing to $D^\prime$. $j\over n$
is Poisson distributed with standard deviation $\sigma(0) =
\sqrt{{p(1-p)}\over n}$. In the continuum one has
$p(k)={\cal N} e^{-{{2\pi}\over{(\mu l )^2 }}k^2}$.
For $\mu l\sim 3$
for pure gauge (ignoring the single plaquette $p^*$ in the continuum)
the decay at  higher $k$'s is sufficiently rapid to ignore all but
$k=0,\pm 1$ as a first approximation for the expected errors.
Hence there is a factor of ${p\over 1-p}$ in the ratio
${N^\prime\over D^\prime}$.
This factor will be distributed with a relative error
$\delta ={1\over{\sqrt{np(1-p)}}}$.
To minimize it we would like
to work with $p\approx
{1\over 2}$. This is indeed the case for $\mu l\approx 3$.

In view of the above considerations, we picked
$\mu l = 3$. The continuum limit is approached by
scaling the lattice coupling $g$ with the lattice spacing via
$g={3\sqrt{\pi}\over L}$. Simulations were performed on lattices
ranging from $L=3$ to $L=10$.
Since $\delta \approx {2\over {\sqrt{n}}}$ is a lower
bound on the relative error in ${N^\prime\over D^\prime}$,
we see that in order to reach an accuracy of
a few percent, we need a few thousand lattices at least. This is
quite acceptable if the lattice sizes $L$ can be kept small.

The continuum result
for the dimensionless quantity ${{<\bar\psi_R \psi_L >}\over {\mu}}$
$\equiv
f(\mu l)$ is known exactly
from the work of Sachs and Wipf [6]: $f(3)\approx 0.236$ and
$f(\infty ) \approx .284$. While we have no reason to try
to extrapolate numerically to infinite physical size,
we like the fact the we are
working on a system that is reasonably large in physical terms. The finite
size effect is seen from the exact solution to be of the order of 20\% and
it is nice that we don't have to eliminate it numerically.

\bigskip\bigskip
\centerline{ \bf Algorithm}
\bigskip
We used very direct and unsophisticated algorithms.
Double precision was used throughout the algorithm. The diagonalization
of $\ham\pm $ was done using Jacobi's method and the needed determinants
evaluating the overlaps were computed using $LU$ decomposition. Since
half of the configurations require the evaluation of $L^2$
determinants the algorithm goes, asymptotically, as $L^8$. In
practice, the larger prefactor in front of the $L^6$ corresponding
to Jacobi diagonalization versus the one corresponding to
determinant computation seems to be important for our sizes and the algorithm
scaled slightly worse than $L^6$. The worst case, $L=10$, took an
average of 5 minutes CPU time per lattice on an 100 MHz $\alpha$--workstation.
No attempts were made to speed up the code. The results of this
paper were obtained using the equivalent of a few hundred hours on such
a machine.

\bigskip\bigskip
\centerline{\bf Results}
\bigskip
Our results
are summarized in the table below
where the numerical results are listed
as a function of $L$.
Errors were estimated in several ways
and all gave consistent answers.
We see that within statistical
errors the correct answer is obtained already at $L=5$, so most
of the computer time we used was not really needed.
We felt, however,
that it was necessary to ascertain that nothing bizarre was happening when
one went closer to the continuum limit.

\def\tablerule{\noalign{\hrule}}

\def\hthree{height3pt&\omit && \omit && \omit &\cr}
\def\hfour{height4pt&\omit && \omit && \omit &\cr}
{\null\hfill
\vbox{\offinterlineskip
\hrule
\halign{\vrule#& \hfil# & \vrule# & \hfil# &
\vrule# & \hfil# &
\vrule# \cr
\hfour
& \ \ $L$  && \ \ $f_L(3)$ && \  \# of confs.
& \cr
\hfour
\tablerule\cr
\hfour
& \ \  3 && \ \ 0.0812(32) && \ \ 5000 & \cr
\hthree
& \ \  4 && \ \ 0.1638(66) && \ \ 5000 & \cr
\hthree
& \ \  5 && \ \ 0.224(11) && \ \ 5000 & \cr
\hthree
& \ \  6 && \ \ 0.223(13) && \ \ 5000 & \cr
\hthree
& \ \  7 && \ \ 0.235(14) && \ \ 6000 & \cr
\hthree
& \ \  8 && \ \ 0.238(12) && \ \ 8173 & \cr
\hthree
& \ \ 10 && \ \ 0.224(18) && \ \ 4919 & \cr
\hthree
\hfour
\tablerule\cr
\hfour
& \ \  $\infty$ && 0.236 && Continuum &\cr
\hthree
\hfour}
\hrule}
\hfill}
\bigskip
\centerline{
{\bf Table:} $ f_L (\mu l) \equiv {L N\over{(\mu l ) D}}$
 for various lattice spacings ${l\over L}$ at $\mu l = 3$.}
\bigskip\bigskip
\centerline{\bf Conclusions and Outlook}
\bigskip
In this paper we have shown that the overlap formalism for chiral
fermions passed a certain ``vector test''.
This test can be used as a check for any
other proposal that regularizes a chiral gauge theory. It is
simple, not very costly to apply and, at present, we believe it
would be a good selection method. It is hard for us to see how a
bilinear method for the regularization of chiral gauge theories
could succeed without passing this vector test. To be sure, let
us stress that the {\it numerical result} we have obtained here
should be obtainable by traditional methods [5]. In these methods
one does not have exact left right factorization. When
one starts from a chiral method this factorization is built in.

The next step would be to study a chiral Schwinger model in the
overlap formalism. Here we do not have exact gauge invariance on
the lattice in the anomaly free case
although this is achieved in the continuum for smooth
gauge fields.  We have to
check whether the Foerster--Nielsen--Ninomya [10] mechanism of
gauge invariance restoration in the continuum limit indeed works
on the lattice
as we conjectured in [2]. Preliminary studies we have done
are encouraging.
Then the road to exhibiting fermion number violation on
the lattice is open.

\bigskip\bigskip
\centerline{\bf Acknowledgements}
\bigskip
This research was supported in part
by the DOE under grant numbers DE-FG02-90ER040542 (RN),
DE-FG05-90ER40559 (HN) and DE-FG02-92ER40699 (PV). Computational facilities at
all three of our institutions were used.

\bigskip\bigskip
\centerline{\bf References}
\bigskip
\item{[1]} G. 't Hooft, Phys. Rev. Lett. 37 (1976).
\item{[2]} R. Narayanan, H. Neuberger,  IASSNS-HEP-94/99, RU-94-83,
hep-th \#9411108,
To appear in Nucl. Phys. B.
\item{[3]} S. Randjbar-Daemi and J. Strathdee,
IC/95/6, hep-lat/9501027; IC/94/401, hep-th/950102;
IC/94/396, hep-th/9412165.
\item{[4]} J. Schwinger, Phys. Rev. 128, 2425 (1962).
\item{[5]} E. Marinari, G. Parisi, C. Rebbi, Nucl. Phys. B190, 734 (1981).
\item{[6]} I. Sachs, A. Wipf, Helv. Phys. Acta 65, 652 (1992).
\item{[7]} H. Joos, Helv. Phys. Acta 63, 670 (1990); H. Joos, S. I. Azakov,
Helv. Phys. Acta 67, 723 (1994); H. Dilger, Nucl. Phys. B434, 321 (1995),
H. Dilger, Phys. Lett. B294, 263 (1992); H. Dilger, DESY 94-145,
hep-lat \# 9408017; H. Dilger, ``The Geometric Schwinger Model
on the Lattice'', thesis, DESY.
\item{[8]} J. Villain, J. Phys. (Paris) 36 (1975) 581.
\item{[9]} R. Narayanan, H. Neuberger, IASSNS-HEP-95/1, RU-94-99,
hep-lat \# 9412104, To appear in Phys. Lett. B.
\item{[10]} D. Foerster, H. B. Nielsen, M. Ninomya, Phys. Lett. 94B,
135 (1980).

\end